\documentclass{article}
\usepackage{gdgspace, emlines2, bezier}
\advance\textheight by -\headsep\advance\textheight by -\headheight
\newcommand{\ketbra}[2]{\vert#1\rangle\langle#2\vert}

\newcommand{\be}{\begin{equation}} 
\newcommand{\ee}{\end{equation}} 
 
\begin{document}
\large
\centerline{\Large Comment on Mohrhoff's ``What Quantum Mechanics
is Trying to Tell Us''}\vskip .3cm
\centerline{March 2, 2001}
\centerline{R. E. Kastner}
\centerline{Department of Philosophy}
\centerline{University of Maryland}
\centerline{College Park, MD 20742}
\vskip 1.5cm

\noindent ABSTRACT. Mohrhoff proposes using 
the Aharonov-Bergmann-Lebowitz (ABL) rule for time-symmetric
``objective'' (meaning non-epistemic) probabilities corresponding to the possible outcomes of 
not-actually-performed measurements between specified pre- and post-selection measurement 
outcomes. 
It is emphasized that the ABL rule was formulated on the assumption that such intervening
measurements are actually made and that it does not necessarily apply
to counterfactual situations. The exact nature of the 
application of the ABL rule considered by Mohrhoff is made explicit
 and is shown to fall short
of his stated counterfactual claim.

\vskip 1cm

Ulrich Mohrhoff\footnotemark[1] distinguishes between ``subjective''
and ``objective'' probabilities in quantum mechanics. This type of distinction
has been made by others\footnotemark[2]; Mohrhoff proposes to modify it
by adding a time-symmetric
aspect in which the relevant ``facts''\footnotemark[3] describing a quantum system
include not just the pre-selection outcome but also the post-selection outcome.
 However, his discussion
of the Aharonov-Bergmann-Lebowitz (``ABL'') rule\footnotemark[4] 
   as related to counterfactual
statements about possible intervening measurements between pre- and
post-selection does not address the numerous objections in
the literature to the counterfactual usage of ABL probabilities.\footnotemark[5]
\footnotemark[6]  Moreover, his usage of the ABL rule implicitly assumes special conditions.
When these special conditions are made explicit, it appears that the actual
claim being made is distinct from, and arguably weaker than, the claim
as stated in [1].

The ABL rule was formulated on the assumption that intervening
measurements at a time $t$, $t_a < t < t_b$, are actually made, and would therefore 
appear to correspond to
what Mohrhoff is calling ``subjective'' probabilities. However,
he wishes to consider time-symmetric {\it objective} probabilities in which 
a possible intervening measurement
of some observable Q is {\it not} actually performed. Various proofs
have been given of the inapplicability of the ABL rule to
this kind of situation. Those proofs demonstrated
that the counterfactual usage of the ABL rule yields consequences
that are inconsistent with quantum theory (see footnote [5], esp. Sharp \& Shanks 1993, Cohen 1995, Miller 1996). 
Vaidman has questioned the validity of such proofs\footnotemark[7] and they have been
defended in Cohen [5, 1998] and Kastner [5]\footnotemark[8]. Thus there is still an active
controversy on this point which Mohrhoff does not address in [1].\footnotemark[9]

Mohrhoff claims that the ABL rule can be applied to statements
such as:

Statement 1: ``If a measurement of observable $Q$ were performed on system $S$ 
between the (actual) 
preparation of the probability measure $\ketbra aa$ at time $t_a$ and 
the (actual) observation of the 
property $\ketbra bb$ at time $t_b$, but no measurement is actually performed
 between $t_a$ and 
$t_b$, then the measurement of $Q$ would yield $q_i$ with probability
 $p\,(q_i|a,b)$.''\footnotemark[10]

\vskip .5cm

Now, if we look at Mohrhoff's application of the ABL rule,
what is actually being done is the following. A possible world {\it j}
is found in which system S happens to have the same pre-
and post-selection outcomes as it did in the actual world ({\it i}).
The ABL rule is then applied, not to the actual world,
but to this possible world {\it j}.\footnotemark[11] Thus Mohrhoff's application
actually supports the following statement, which differs
significantly from Statement (1):

Statement 2: ``In the possible world {\it j} in which observable Q is
measured and system S yields 
outcomes $a$ and $b$ at times $t_a$ and $t_b$ respectively, 
the probability of obtaining result $q_k$ at 
time $t$ is given by $P_{ABL}(q_k|a,b)$.'' 
\vskip .5cm

	It should be noted that the possible world {\it j}
referred to in Statement 2---the one in which system S happens to
end up with the same pre- and post-selection results as in
the actual world---
is only one member of a set W of worlds which differ from the actual
world in that observable Q {\it is} measured.
The other worlds in this set are worlds in which observable Q is measured 
at time $t$ and
the pre- and/or post-selection outcomes are {\it not} the same
as in the actual world. 

To sum up the situation so far: In Mohrhoff's counterfactual claim, 
the ABL rule is applied not to the actual world (in which no measurement 
was performed at time $t$), but rather to a possible
world in which the measurement at time $t$ is performed
and the quantum system has the same pre- and post-selection
outcomes as in the actual world. The result thus obtained is claimed to apply to
the actual world inasmuch as it is claimed to give an answer to a question
about how our world might have been different between
the times $t_a$ and $t_b$ (with their associated outcomes), were a certain measurement
made (i.e., it is claimed to provide a basis for the truth of Statement 1).
Let us call this kind of claim a ``counterfactual\dag,''
 where the ``\dag''
signifies that it is a new type of counterfactual claim which
may or may not be immune to the
type of objections previously raised in the literature. Our task,
then, is to analyse this counterfactual\dag\ and determine whether
it avoids the conclusion
of proofs such as that of Sharp \& Shanks [5].\footnotemark[12]

Let us first clarify what is involved in considering hypothetical
situations involving measurements that were not actually made.
Consider the following question from the viewpoint of an experimenter---
a physicist to whom we will henceforth refer as 
``Dr. X''---
who preselects particles in state $|a\rangle$ at $t_a$ and post-selects
particles in state $|b\rangle$ at $t_b$.

Dr. X asks himself:
 ``In general, what would have happened if I had made a measurement at time $t$
that I did not, in fact make? How might the data of my experiment
change?''

 One part of the answer to this question is that particles that were found
to be in state $|b\rangle$ at $t_b$ might not have ended up in the same state.
\footnotemark[13]

Therefore, from Dr. X's point of view, the appropriate 
and correct counterfactual
statement of the ABL rule would be worded as follows:

Statement 1$'$: ``Consider system S having pre- and post-selection results $a$ and $b$ at times
$t_a$ and $t_b$ when a measurement of observable $Q$ was not performed.
If a measurement of observable $Q$ had 
been performed at time
$t,  t_a < t < t_b$ on S, {\it and if S had the same pre- and post-selection
outcomes as above},
 outcome $q_k$ would have resulted with probability $P_{ABL}(q_k|a,b)$,''\vskip .5cm

Now, for Dr. X the second, italicized ``if'' clause is a big ``if,'' in view of his
question and answer above. Acknowledging this second ``if'' takes into account
that the background conditions that must hold 
in order to apply the ABL rule counterfactually to system S---namely, that it must
have the same pre- and post-selection results as in the actual world---are not
guaranteed to hold if the measurement is actually performed. (See footnote [13].)

In the absence of the caveat of the second ``if,'' which considerably
weakens the counterfactual claim, such
background conditions have to satisfy the following
requirement, which is itself a counterfactual statement:
\vskip .3cm
Requirement C: If a measurement of observable Q
had been performed, system S would (with certainty) have been pre- and post-selected
with outcomes $a$ and $b$ as in the actual world.
\vskip .3cm
This requirement is often referred to as ``cotenability'': if the
necessary background conditions are not {\it cotenable} with the
antecedent (i.e., the measurement of Q), then a counterfactual
statement such as Statement 1, which crucially depends on the stability of 
those background conditions, fails. 

Now, it is obvious that in the pre- and post-selection situation
considered by Mohrhoff, Requirement C is not satisfied. As Dr. X observes
above, if he had in fact measured Q, then system S might not
have been post-selected. However, Mohrhoff argues that cotenability
is not an issue for his new counterfactual\dag\ ; all he
requires is that a possible world {\it j} exists in which
he can apply the ABL rule, with pre- and post-selection
results $a$ and $b$, to system S. Therefore his argument is essentially
the following: to make a counterfactual claim like that of
Statement (1), all one need do is to find a possible world
in which required background conditions happen to hold,
apply the ABL rule to the counterpart of system S in that
possible world, and then claim that the result applies
to the actual system S. 

A diagram may help to make clear the exact nature of Mohrhoff's
counterfactual\dag\ claim.\footnotemark[14] It is argued that
one should take into account that the outcome of the final measurement
was the value $b$. This outcome should be viewed as ``fixed,''
the idea being that the final conditions should have the same status
as the initial conditions; in other words, one assumes a two-valued temporal
boundary condition. This is equivalent to assuming that the
appropriate probability to be assigned to the event $b$ at $t_2$
is unity, the {\it posterior} probability of outcome $b$ 
(since it actually happened). That is, 
questions about what might have happened other than outcome $b$,
are now viewed as irrelevant, since the applicable probabilities
for such questions are {\it prior} probabilities and as such do
not take into account all relevant facts.

 Now, one considers a system of possible
worlds with the following characteristics (refer to Figure 1):\vskip .5cm
\special{em:linewidth 0.4pt}
\unitlength 1.00mm
\linethickness{0.4pt}
\begin{picture}(131.33,134.33)
\emline{30.67}{99.67}{1}{33.90}{101.87}{2}
\emline{33.90}{101.87}{3}{37.12}{103.92}{4}
\emline{37.12}{103.92}{5}{40.33}{105.82}{6}
\emline{40.33}{105.82}{7}{43.53}{107.58}{8}
\emline{43.53}{107.58}{9}{46.73}{109.19}{10}
\emline{46.73}{109.19}{11}{49.92}{110.65}{12}
\emline{49.92}{110.65}{13}{53.10}{111.97}{14}
\emline{53.10}{111.97}{15}{56.27}{113.14}{16}
\emline{56.27}{113.14}{17}{59.44}{114.17}{18}
\emline{59.44}{114.17}{19}{62.59}{115.05}{20}
\emline{62.59}{115.05}{21}{65.74}{115.78}{22}
\emline{65.74}{115.78}{23}{68.88}{116.36}{24}
\emline{68.88}{116.36}{25}{72.02}{116.80}{26}
\emline{72.02}{116.80}{27}{75.14}{117.09}{28}
\emline{75.14}{117.09}{29}{78.26}{117.24}{30}
\emline{78.26}{117.24}{31}{81.37}{117.23}{32}
\emline{81.37}{117.23}{33}{84.47}{117.09}{34}
\emline{84.47}{117.09}{35}{87.57}{116.79}{36}
\emline{87.57}{116.79}{37}{90.65}{116.35}{38}
\emline{90.65}{116.35}{39}{93.73}{115.76}{40}
\emline{93.73}{115.76}{41}{96.80}{115.03}{42}
\emline{96.80}{115.03}{43}{99.87}{114.15}{44}
\emline{99.87}{114.15}{45}{102.92}{113.12}{46}
\emline{102.92}{113.12}{47}{105.97}{111.94}{48}
\emline{105.97}{111.94}{49}{109.01}{110.62}{50}
\emline{109.01}{110.62}{51}{112.04}{109.15}{52}
\emline{112.04}{109.15}{53}{115.06}{107.54}{54}
\emline{115.06}{107.54}{55}{118.08}{105.78}{56}
\emline{118.08}{105.78}{57}{121.09}{103.87}{58}
\emline{121.09}{103.87}{59}{125.67}{100.67}{60}
\emline{33.33}{97.00}{61}{36.77}{97.98}{62}
\emline{36.77}{97.98}{63}{40.20}{98.88}{64}
\emline{40.20}{98.88}{65}{43.62}{99.72}{66}
\emline{43.62}{99.72}{67}{47.02}{100.49}{68}
\emline{47.02}{100.49}{69}{50.41}{101.19}{70}
\emline{50.41}{101.19}{71}{53.78}{101.82}{72}
\emline{53.78}{101.82}{73}{57.14}{102.38}{74}
\emline{57.14}{102.38}{75}{60.49}{102.87}{76}
\emline{60.49}{102.87}{77}{63.83}{103.29}{78}
\emline{63.83}{103.29}{79}{67.15}{103.64}{80}
\emline{67.15}{103.64}{81}{70.45}{103.93}{82}
\emline{70.45}{103.93}{83}{73.75}{104.14}{84}
\emline{73.75}{104.14}{85}{77.02}{104.29}{86}
\emline{77.02}{104.29}{87}{80.29}{104.36}{88}
\emline{80.29}{104.36}{89}{83.54}{104.37}{90}
\emline{83.54}{104.37}{91}{86.78}{104.31}{92}
\emline{86.78}{104.31}{93}{90.00}{104.17}{94}
\emline{90.00}{104.17}{95}{93.22}{103.97}{96}
\emline{93.22}{103.97}{97}{96.41}{103.70}{98}
\emline{96.41}{103.70}{99}{99.60}{103.36}{100}
\emline{99.60}{103.36}{101}{102.77}{102.95}{102}
\emline{102.77}{102.95}{103}{105.92}{102.48}{104}
\emline{105.92}{102.48}{105}{109.06}{101.93}{106}
\emline{109.06}{101.93}{107}{112.19}{101.31}{108}
\emline{112.19}{101.31}{109}{115.31}{100.63}{110}
\emline{115.31}{100.63}{111}{118.41}{99.87}{112}
\emline{118.41}{99.87}{113}{121.67}{99.00}{114}
\emline{33.33}{93.33}{115}{37.10}{91.64}{116}
\emline{37.10}{91.64}{117}{40.81}{90.07}{118}
\emline{40.81}{90.07}{119}{44.48}{88.62}{120}
\emline{44.48}{88.62}{121}{48.10}{87.30}{122}
\emline{48.10}{87.30}{123}{51.67}{86.11}{124}
\emline{51.67}{86.11}{125}{55.20}{85.03}{126}
\emline{55.20}{85.03}{127}{58.67}{84.09}{128}
\emline{58.67}{84.09}{129}{62.10}{83.26}{130}
\emline{62.10}{83.26}{131}{65.48}{82.56}{132}
\emline{65.48}{82.56}{133}{68.81}{81.98}{134}
\emline{68.81}{81.98}{135}{72.09}{81.53}{136}
\emline{72.09}{81.53}{137}{75.32}{81.20}{138}
\emline{75.32}{81.20}{139}{78.51}{81.00}{140}
\emline{78.51}{81.00}{141}{81.64}{80.92}{142}
\emline{81.64}{80.92}{143}{84.73}{80.96}{144}
\emline{84.73}{80.96}{145}{87.77}{81.13}{146}
\emline{87.77}{81.13}{147}{90.76}{81.42}{148}
\emline{90.76}{81.42}{149}{93.71}{81.83}{150}
\emline{93.71}{81.83}{151}{96.60}{82.37}{152}
\emline{96.60}{82.37}{153}{99.45}{83.03}{154}
\emline{99.45}{83.03}{155}{102.24}{83.82}{156}
\emline{102.24}{83.82}{157}{104.99}{84.73}{158}
\emline{104.99}{84.73}{159}{107.69}{85.77}{160}
\emline{107.69}{85.77}{161}{110.34}{86.93}{162}
\emline{110.34}{86.93}{163}{112.95}{88.21}{164}
\emline{112.95}{88.21}{165}{115.50}{89.62}{166}
\emline{115.50}{89.62}{167}{118.01}{91.15}{168}
\emline{118.01}{91.15}{169}{121.67}{93.67}{170}
\emline{33.67}{95.33}{171}{41.07}{94.31}{172}
\emline{41.07}{94.31}{173}{44.72}{93.86}{174}
\emline{44.72}{93.86}{175}{48.34}{93.46}{176}
\emline{48.34}{93.46}{177}{51.92}{93.09}{178}
\emline{51.92}{93.09}{179}{55.47}{92.77}{180}
\emline{55.47}{92.77}{181}{58.99}{92.49}{182}
\emline{58.99}{92.49}{183}{62.48}{92.25}{184}
\emline{62.48}{92.25}{185}{65.93}{92.05}{186}
\emline{65.93}{92.05}{187}{69.35}{91.89}{188}
\emline{69.35}{91.89}{189}{72.74}{91.77}{190}
\emline{72.74}{91.77}{191}{76.09}{91.70}{192}
\emline{76.09}{91.70}{193}{79.41}{91.67}{194}
\emline{79.41}{91.67}{195}{82.70}{91.68}{196}
\emline{82.70}{91.68}{197}{85.95}{91.72}{198}
\emline{85.95}{91.72}{199}{89.18}{91.82}{200}
\emline{89.18}{91.82}{201}{92.37}{91.95}{202}
\emline{92.37}{91.95}{203}{95.52}{92.12}{204}
\emline{95.52}{92.12}{205}{98.64}{92.34}{206}
\emline{98.64}{92.34}{207}{101.73}{92.59}{208}
\emline{101.73}{92.59}{209}{104.79}{92.89}{210}
\emline{104.79}{92.89}{211}{107.82}{93.23}{212}
\emline{107.82}{93.23}{213}{110.81}{93.61}{214}
\emline{110.81}{93.61}{215}{113.77}{94.03}{216}
\emline{113.77}{94.03}{217}{116.69}{94.50}{218}
\emline{116.69}{94.50}{219}{121.33}{95.33}{220}
\put(66.33,118.33){\makebox(0,0)[cc]{$\{Q_1\}$}}
\put(72.00,106.00){\makebox(0,0)[cc]{$\{Q_2\}$}}
\put(69.67,94.00){\makebox(0,0)[cc]{$\{Q_3\}$}}
\put(72.00,83.67){\makebox(0,0)[cc]{$\{Q_4\}$}}
\put(62.33,69.00){\makebox(0,0)[cc]{$\{Q_5\}$}}
\put(28.33,87.33){\makebox(0,0)[cc]{$t_1$}}
\put(126.00,87.00){\makebox(0,0)[cc]{$t_2$}}
\put(24.67,91.00){\framebox(7.33,7.33)[cc]{$a$}}
\put(78.67,39.67){\makebox(0,0)[cc]
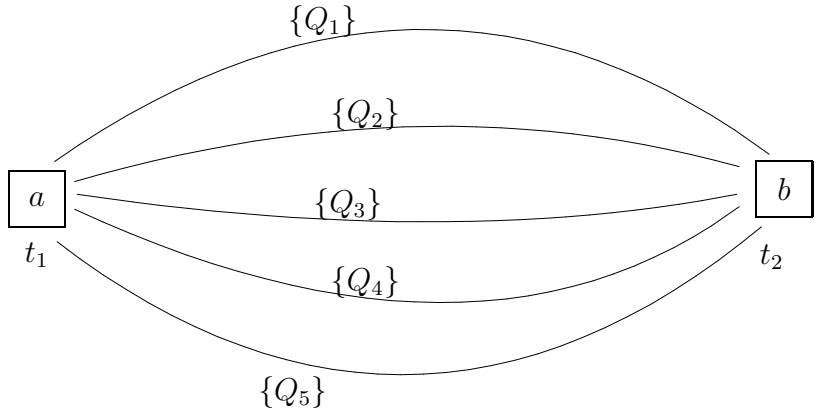
{Figure 1. The system of possible worlds assumed in Mohrhoff's ABL counterfactual\dag.}}
\emline{31.00}{89.00}{221}{34.06}{86.73}{222}
\emline{34.06}{86.73}{223}{37.12}{84.61}{224}
\emline{37.12}{84.61}{225}{40.18}{82.65}{226}
\emline{40.18}{82.65}{227}{43.23}{80.85}{228}
\emline{43.23}{80.85}{229}{46.28}{79.20}{230}
\emline{46.28}{79.20}{231}{49.33}{77.71}{232}
\emline{49.33}{77.71}{233}{52.38}{76.38}{234}
\emline{52.38}{76.38}{235}{55.43}{75.20}{236}
\emline{55.43}{75.20}{237}{58.47}{74.18}{238}
\emline{58.47}{74.18}{239}{61.51}{73.31}{240}
\emline{61.51}{73.31}{241}{64.55}{72.60}{242}
\emline{64.55}{72.60}{243}{67.59}{72.05}{244}
\emline{67.59}{72.05}{245}{70.62}{71.65}{246}
\emline{70.62}{71.65}{247}{73.66}{71.41}{248}
\emline{73.66}{71.41}{249}{76.69}{71.32}{250}
\emline{76.69}{71.32}{251}{79.72}{71.39}{252}
\emline{79.72}{71.39}{253}{82.75}{71.62}{254}
\emline{82.75}{71.62}{255}{85.77}{72.00}{256}
\emline{85.77}{72.00}{257}{88.79}{72.54}{258}
\emline{88.79}{72.54}{259}{91.82}{73.24}{260}
\emline{91.82}{73.24}{261}{94.83}{74.09}{262}
\emline{94.83}{74.09}{263}{97.85}{75.10}{264}
\emline{97.85}{75.10}{265}{100.87}{76.26}{266}
\emline{100.87}{76.26}{267}{103.88}{77.58}{268}
\emline{103.88}{77.58}{269}{106.89}{79.06}{270}
\emline{106.89}{79.06}{271}{109.90}{80.69}{272}
\emline{109.90}{80.69}{273}{112.91}{82.48}{274}
\emline{112.91}{82.48}{275}{115.91}{84.43}{276}
\emline{115.91}{84.43}{277}{118.91}{86.53}{278}
\emline{118.91}{86.53}{279}{121.91}{88.79}{280}
\emline{121.91}{88.79}{281}{124.67}{91.00}{282}
\put(124.00,92.33){\framebox(7.33,7.33)[cc]{$b$}}
\end{picture}

Each of the curved lines represents a set of possible worlds $\{Q_j\}$
in which a measurement of the observable $Q_j$ is performed at
time $t$. 
The individual member worlds of each set $\{Q_j\}$ 
are identified with the possible outcomes of measurements of $Q_j$.
(Thus a particular set of worlds $\{Q_j\}$ here corresponds
to the possible world $\it j$ referred to above, with additional
structure.) 
Within this proposed construct, the 
ABL probability is assumed to give the {\it subjective}
probability applying to an observer Dr. X\dag\  associated with the
set of worlds $\{Q_j\}$ in which the measurement of $Q_j$ is 
performed.\footnotemark[15]
In Mohrhoff's counterfactual\dag\ , this probability is then held to 
apply (objectively) to a different world---the world
of Dr. X---in that it is claimed to give an answer to the question
corresponding to Statement (1), i.e.: ``If (contrary to fact)
a measurement of $Q_j$ had
been performed between the actual outcomes at
$t_a$ and $t_b$, what would be the probability that Dr. X would
find outcome ${q_j}_k$ (one of the eigenvalues of $Q_j$)?''

However, such a claim would seem to conflate two distinctly different 
perspectives
or frames of reference. In being invoked in Statement (1), the ABL 
rule is being applied counterfactually to Dr. X's world, since it is from that
frame of reference that the claim is being made. Yet when called upon
to justify his counterfactual\dag\ , Mohrhoff argues from the 
non-counterfactual perspective of Dr. X\dag\ . 

For example, Mohrhoff's objection to the Sharp \& Shanks proof, like
Vaidman's, consists in demanding that the ``counterfactual'' measurement 
be considered as performed, in contrast to the proof which assumes
that, in accordance with Dr. X's perspective, the measurement is {\it not}
performed. Thus, according to Mohrhoff, the calculation should be
considered as applying to the world(s) of Dr. X\dag\  rather than 
to the world of Dr. X.
But Dr. X\dag\ 's perspective is markedly different
from that of Dr. X, and it is Dr. X who is making the counterfactual claim 
(Statement 1).  The proof applies to Dr. X and his claim, 
not to someone who has actually performed the measurement.
One does not refute a proof by arguing that
it does not apply to a frame
of reference for which it was not intended. Thus Mohrhoff's counterfactual\dag\ 
formulation, which postulates a hypothetical system of
worlds in which {\it non}-counterfactual applications of the ABL rule 
are held to apply
to those possible worlds, fails to evade the conclusion of the proofs
which address the {\it actual} world in which no such measurement is performed.

To conclude, it has been argued that Mohrhoff's application of the ABL
rule, which depends on a specially chosen system of possible worlds,
fails to support his stated counterfactual claim based on that rule. 
Moreover, his objection to the proofs demonstrating
the nonvalidity of the counterfactual usage of the ABL rule fails
to refute those proofs, because it does not apply
to the actual world addressed by the proofs.

Finally, it should be noted that none of the arguments in this Comment 
presuppose any time-asymmetry assumptions. 
 Nor are any underlying metaphysical assumptions about
time ``flow'' or subjective, ``classical'' experience invoked or required.
The present
author fully agrees with the basic result (relevant to this
discussion) originally obtained by ABL, which
was the following: ``...during the time interval between two noncommuting
observations, we may assign to a system the quantum state corresponding to
the observation that follows with as much justification as we assign,
ordinarily, the state corresponding to the preceding measurement.''\footnotemark[16]

The above quoted statement considers {\it either}
temporal direction as being equally valid as far as quantum theory
is concerned (where here, ``temporal direction'' simply means
whether one regards the parameter $t$ as increasing or decreasing
in the applicable laws). However, it does not consider {\it combining
both temporal directions} as is implied in a counterfactual
usage of the ABL rule. I.e., nothing in the ABL paper suggests holding
fixed {\it both} pre- and post-selection states while considering not-actually
-performed
measurements during that time interval.\footnotemark[17]
\vskip 1cm
Acknowledgements.  \newline
I would like to thank Ulrich Mohrhoff for an interesting
exchange of views.
\newpage

\footnotetext[1] {\normalsize U. Mohrhoff. `What Quantum Mechanics
is Trying to Tell Us,'
{\it American Journal of Physics 68}, 728-745 (2000).}

\footnotetext[2] {\normalsize See, for example, R.I.G. Hughes. {\it The Structure and Interpretation
of Quantum Mechanics} (Harvard University Press, Cambridge, MA, 1989, p. 218.; A. Shimony.
``Search for a worldview which can accomodate our knowledge of microphysics,''
in J. Cushing and E. McMullin, eds., {\it Philosophical Consequences of Quantum Theory}
(University of Notre Dame Press, Notre Dame, 1989), p. 27; A. Shimony,
{\it Search for a Naturatistic World View, Vol. II} (Cambridge University Press,
New York, NY, 1993), pp. 141-2. Mohrhoff defines subjective probabilities as 
applying only in cases
in which measurements have been made and an observer is ignorant
of the result of the measurement, which differs slightly from Hughes'
use of the term (see note [3]).} 
\footnotetext[3] {\normalsize Mohrhoff applies the term ``fact'' to measurement
outcomes only (whether known or unknown), as opposed to
possessed properties independent of measurement (which he denies).}
\footnotetext[4] {\normalsize
Y. Aharonov, P. G. Bergmann, and J. L. Lebowitz. 
`Time Symmetry in the Quantum Process of Measurement,' 
{\it Physical Review B 134}, 1410-16 (1964).}
\footnotetext[5] {\normalsize See, for instance, 
J. Bub and H. Brown. `Curious Properties of Quantum Ensembles
Which Have Been Both Preselected and Post-Selected,' {\it Physical Review Letters 56},
2337-2340 (1986); W. Sharp and N. Shanks. `The Rise and Fall of 
Time-Symmetrized 
Quantum Mechanics,' {\it Philosophy of Science 60}, 
488-499 (1993); O. Cohen. `Pre- and postselected quantum systems, 
counterfactual 
measurements, and consistent histories, '
{\it Physical Review A 51}, 4373-4380 (1995); `Reply to
``Validity of the Aharonov-Bergmann-Lebowitz Rule'', '
{\it Physical Review A 57}, 2254-2255 (1998);
D. J. Miller. `Realism and Time Symmetry in 
Quantum Mechanics,' {\it Physics Letters A 222}, 31-36 (1996); R. E. Kastner.
 `Time-Symmetrised Quantum Theory,
Counterfactuals, and ``Advanced Action,''' 
{\it Studies in History and Philosophy of Modern Physics 30},
 237-259 (1999);  `The Three-Box Paradox and Other Reasons
to Reject the Counterfactual Usage of the ABL Rule,'
{\it Foundations of Physics 29}, 851-863 (1999); `TSQT ``Elements of Possibility''?,'
{\it Studies in History and Philosophy of Modern Physics 30},
399-402 (1999).}
\footnotetext[6] {\normalsize As it happens, Mohrhoff's specific example of a counterfactual
use of the ABL rule corresponds to a special case in which
that use is valid (in the strong sense of Statement 1). This is an example in which a particle is
pre- and post-selected with outcomes $a$ and $b$ corresponding to noncommuting
observables A and B, and counterfactual measurements of either
A or B are considered at time $t$. (The validity of a counterfactual
usage of the ABL rule in cases like this has been shown in 
detail in Kastner [5]  and in Cohen [5].) But this is a special
case and, as has been discussed at length in the literature, 
``would''-type counterfactual
uses of the ABL rule are generally invalid.}
\footnotetext[7] {\normalsize L. Vaidman. `Validity of the Aharonov-Bergmann-Lebowitz Rule,' 
{\it Physical Review A 57}, 2251-2253 (1998).}
\footnotetext[8] {\normalsize In his reply to this
Comment, Mohrhoff rejects my
defense of the Sharp and Shanks proof on the basis that I 
allegedly assume the reality of ensembles
in situations concerning only one particle.
This remark misunderstands my use of ensembles. The
term ``ensemble'' as I am using it denotes not a real
collection of particles, but rather a
{\it conceptual} ensemble in the sense of statistical mechanics
(cf. R. K. Pathria. {\it Statistical Mechanics} (Pergamon Press, Oxford, 1972),
p. 4).}
\footnotetext[9]{\normalsize In his reply to this Comment, Mohrhoff essentially repeats 
the objection to the Sharp and Shanks proof previously
given by Vaidman[7]. In this attempt to
refute the proofs of Sharp and Shanks and others
(which all have essentially the same structure), Vaidman and Mohrhoff
simply assume that the ``counterfactual'' measurement
{\it is performed} in all expressions employed in the proof. Then, of course, there
can be no inconsistency with the predictions of quantum mechanics. But this is
no refutation, for it explicitly {\it assumes as true that which is
manifestly false}: namely, that the ``counterfactual'' measurement
is actually made. This procedure is then justified by claiming
that the computation applies not to the actual world but
to a specially chosen possible world. If such a procedure were to be allowed,
then one could argue for something manifestly false in the actual
world merely by finding a specially chosen possible world in which it is true.}
\footnotetext[10] {\normalsize `Objective probabilities, quantum counterfactuals,
and the ABL rule: Apropos of Kastner's comment,' quant-ph/0006116 (2000) 
(a preprint version of Mohrhoff's reply to the present Comment).}
\footnotetext[11] {\normalsize U. Mohrhoff, quant-ph/0006116, p. 4:
 ``If we think of the measurement of $Q$ as taking
place in a possible world, we consider a world in which all the
relevant facts are exactly as they are in the actual world, except
that in this possible world there is one additional relevant
fact indicating the value possessed by $Q$ at a time between
$t_a$ and $t_b$.''}
\footnotetext[12]{\normalsize Sharp and Shanks (1993)
 consider an ensemble of spin-$1\over2$
 particles  prepared at time $t_1$ in the state 
$\vert a_1\rangle$
 (read as `spin up along direction {\bf a}'). They then
 assume that this ensemble is subjected to a final post-
selection spin measurement at time $t_2$ along direction
{\bf b} (i.e., the observable $\sigma_b$ is measured).
This measurement yields two subensembles $E_i, i=1,2$
corresponding to results spin up or spin down along direction
{\bf b}.  The weight of 
each subensemble $E_i$ is given by
 $\vert \langle b_i\vert a_1\rangle \vert ^2$.
 
	Now they consider each subensemble individually, 
asking the counterfactual question: If we had measured the
 spin of these particles along direction {\bf c}
 (i.e., observable $\sigma_c$)  at a time $t$ between \  $t_1$ \ and \ $t_2$, what
 would have been the probability for outcome $c_1$? They use
the ABL rule to calculate the probability of  outcome $c_1$
 for each subensemble $E_i$ for such a counterfactual
 measurement. They then show that the total probability of 
outcome $c_1$ derived from the above calculation, taking into
 account  the weights of the two subensembles $E_i$, in
 general disagrees with the quantum mechanical probability,
 which is given simply by
 $\vert \langle c_i\vert a_1\rangle \vert ^2$.}
\footnotetext[13] {\normalsize One can, of course, deny this statement if
one assumes fatalism (i.e., everything that happens
{\it must} happen). But then it must
also be assumed that there is no possibility of a ``counterfactual''
measurement at time $t$, since it is a recordable matter of fact that no such
measurement occurred, and according to fatalism, that documented absence of
a measurement is also a fact that {\it must} happen.}
\footnotetext[14]{\normalsize Mohrhoff has confirmed in a private correspondence
that the diagram discussed
herein correctly illustrates his proposed possible world structure.}
\footnotetext[15] {\normalsize However, this assumption can
be disputed, since under the given construct (which assumes
that outcome $b$ definitely occurs at $t_2$), the conditional probabilities
$P(b|q_{j{_k}})$ (where $q_{j{_k}}$ is an eigenvalue of the associated
observable $Q_j$) are unity, rather than the standard quantum mechanical
conditional probabilities as assumed in the ABL rule.}
\footnotetext[16] {\normalsize ABL (1964), abstract.}
\footnotetext[17] {\normalsize ABL (1964) do say ``We shall now
consider an ensemble of systems whose initial and final states are fixed to
correspond to the particular eigenvalues $a$ and $b$, respectively;
we ask for the probability that the outcome of the intervening
measurements are $d_j, ... d_n$, respectively.'' (p. B1412) But those outcomes
correspond to {\it actually performed} measurements.}

\end{document}